\documentclass[conference]{IEEEtran}
\IEEEoverridecommandlockouts
\usepackage{cite}
\usepackage{amsmath,amssymb,amsfonts,mathtools}
\usepackage{algorithmic}
\usepackage{graphicx}
\usepackage{textcomp}
\usepackage{comment}
\usepackage{xcolor}

\usepackage{amsthm}
\theoremstyle{plain}

\usepackage{tabularx}  
\usepackage{multirow} 

\usepackage{array} 
\newcolumntype{L}{>{\raggedright\arraybackslash}X}

\usepackage{booktabs} % \toprule \midrule \bottomrule

\usepackage{hyperref} % enables \texorpdfstring and hyperlinks
\usepackage[capitalise,nameinlink]{cleveref} % smart refs: \Cref{eq:...}

\def\BibTeX{{\rm B\kern-.05em{\sc i\kern-.025em b}\kern-.08em
    T\kern-.1667em\lower.7ex\hbox{E}\kern-.125emX}}
\begin{document}

\title{OntoTune: Ontology-Driven Learning for Query Optimization with Convolutional Models\\
%\title{OntoTune: Query Optimization via Ontology-Embedded Graph Neural Networks\\
}

\author{\IEEEauthorblockN{1\textsuperscript{st} Songhui Yue}
\IEEEauthorblockA{
%\textit{Computer Science Department} \\
\textit{Charleston Southern University}\\
syue@csunive.edu}
\and
\IEEEauthorblockN{2\textsuperscript{nd} Yang Shao}
\IEEEauthorblockA{
%\textit{Computer Science Department} \\
\textit{Charleston Southern University}\\
yshao@student.csuniv.edu}
\and
\IEEEauthorblockN{3\textsuperscript{th} Sean Hayes}
\IEEEauthorblockA{
%\textit{Computer Science Department} \\
\textit{Charleston Southern University}\\
shayes@csuniv.edu}

}

\maketitle

\begin{figure*}[htbp]
	\centering
	\includegraphics[width=0.9\linewidth]{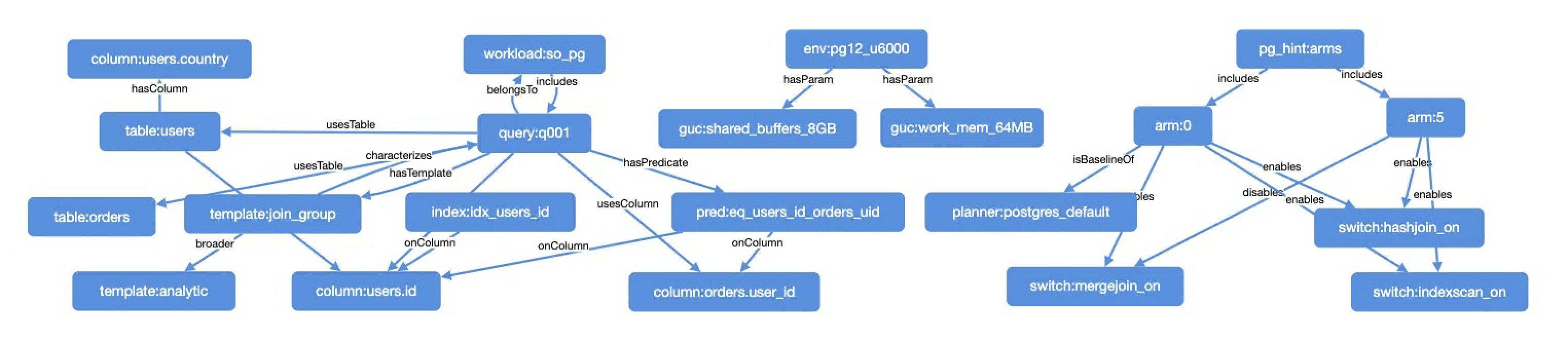}
	\caption{A Partial Knowledge Graph for DB Query Optimization (with Ontology and Instance Examples)}
	\label{onto_db}
\end{figure*}

\begin{abstract}
Query optimization has been studied using machine learning, reinforcement learning, and, more recently, graph-based convolutional networks. Ontology, as a structured, information-rich knowledge representation, can provide context, particularly in learning problems. This paper presents OntoTune, an ontology-based platform for enhancing learning for query optimization. By connecting SQL queries, database metadata, and statistics, the ontology developed in this research is promising in capturing relationships and important determinants of query performance. This research also develops a method to embed ontologies while preserving as much of the relationships and key information as possible, before feeding it into learning algorithms such as tree-based and graph-based convolutional networks. A case study shows how OntoTune’s ontology-driven learning delivers performance gains compared with database system default query execution.
\end{abstract}


\section{Introduction}
Database query optimization has been studied for decades, starting from fixed heuristic models, to learning models \cite{zhu2023lero, gadde2024intelligent}, and the latter highly relies on deep reinforcement learning models \cite{ortiz2018learning}, and other deep learning models \cite{yang2022machine}, such as a transformer model \cite{zhao2022queryformer}. Those studies are performed on query datasets that are composed of different types \cite{heitz2019join, marcus2021bao}. These queries often arise from template families (e.g., particular table-usage patterns or the presence of DISTINCT/COUNT). The database workload contains the tables and attributes, and their related information includes table and attribute schemas, index availability, data distributions and skew, and the quality of cardinality estimates. All these types of information, along with the execution environment, when organized into features, can determine the execution speed of a query set in a particular database system, such as PostgreSQL, a widely used open-source system.

Modern workloads such as the Jobs/Join-Order Benchmark (JOB) \cite{leis2015good} and StackOverflow (so\_pg) \cite{marcus_stack_sopg} expose a persistent challenge: long-tail queries and environment-sensitive behavior that fixed-cost models fail to capture \cite{marcus2021bao}. Learning-based approaches help, but they can be brittle: outcomes depend on batch size, random seed, hot vs.\ cold start, and on hardware, statistics, and DBMS configuration (GUCs) \cite{giannakouris2025lambda}. From a modeling standpoint, ML predictors exhibit high variance across training sets \cite{ding2019ai} and suffer when deployed under distribution shift, yielding large errors on data that differ from the training distribution \cite{ding2019ai,ma2020active}. These factors make experiments unstable, hard to reproduce, and difficult to explain.

In this study, we propose OntoTune, an ontology-driven learning platform for query optimization. Our view in this study is simple: an ontology and a concrete knowledge graph (KG) provide the missing semantic and provenance layers for learned optimization. We name the objects and relations that matter to plan selection, and we record runs with their full context. Queries, plan trees and operators, planner arms (i.e., bundles of PostgreSQL GUC switches), runtime metrics, environments, sampling policies, and model predictions/rewards all become first-class, queryable entities. The ontology (TBox) \cite{ben2021novel} gives the schema, and the KG (ABox) \cite{simsek2023knowledge} stores instances. 

Ontotune is a PostgreSQL-based prototype. The system follows a similar implementation architecture to Bao (a learned component that sits on top of the native optimizer) \cite{marcus2021bao}, to which we add a query-related metadata and system configuration extraction module via the PG extension, and a workflow on the training side that utilizes extracted data as feature providers. By preserving the query metadata, DB statistics, plan trees and operators, runtime and buffer metrics, and relevant Grand Unified Configurations (GUCs) (such as \texttt{enable\_*} and \texttt{work\_mem}), experiments are traceable and comparable across configurations.

For the utilization of the ontology and KG, we adopt a CNN-based predictor \cite{ma2022understanding} trained on an ontology-derived feature matrix that fuses SQL, plan, and context information. On the datasets considered, the CNN is our strongest variant. For online arm selection (bandit-style reinforcement learning), we apply a 1-complement reward transform to reduce the autoselection of seldom-used arms. Because the representation lives at the ontology/KG layer, tree- and graph-based models (TreeConv \cite{marcus2021bao}, GCN \cite{zhang2019graph}, GAT \cite{velivckovic2017graph}) can plug into the same data pipeline without re-engineering; we treat them as forward-looking components for future exploration. Case studies illustrate both regimes: when the learned policy outperforms the PostgreSQL baseline and when gains disappear (e.g., under cold starts or highly exploratory settings).

We format the database and query context as an ontology/knowledge graph (KG) in this application for four reasons: First, interoperability \cite{lee2023relation}: a minimal, task-driven vocabulary avoids over-engineering while standardizing names and relations. Second, explanation \cite{rajabi2024knowledge}: the KG ties each run to its query, arm, plan, and metrics, making case analysis straightforward. Third, reproducibility: environment/GUC snapshots, plan fingerprints, and sampling choices (hot/cold start, batch regimes, exploration) are part of the data, not side notes. Figure~\ref{onto_db} sketches the overall design: the ontology defines the core classes; the KG records per-run evidence and links among them. Fourthly, the ontology+KG can provide structured representations that feed into AI models, such as LLMs \cite{giannakouris2025lambda}, to improve context understanding for further query optimization and transfer learning.

The contributions of this study are summarized as follows:
\begin{itemize}
	\item A minimal ontology (TBox) and a practical knowledge graph (ABox) for learning-based query optimization, covering queries, plans, operators, arms (GUC bundles), environments, metrics, predictions, and rewards.
	\item OntoTune: a PostgreSQL-based platform that mirrors prior learned-optimizer integration but adds end-to-end provenance capability: plan extraction, metric collection, and explicit snapshots of environment and GUC settings.
	\item The case study combines CNN into the reinforcement learning process through designing a reward-cost complement transform trick to avoid the autoselection of seldomly-used arms between different batches. CNNs provide our strongest results so far on selected datasets.
\end{itemize}

The subsequent sections are structured as follows: Section~\ref{sec:methodology} introduces the ontology and KG schema and how we populate and embed them into feature matrices. Section~\ref{sec:implementation} describes the implementation of the learning pipeline. Section~\ref{sec:experiments} presents case studies and evaluation under fixed configurations. Section~\ref{sec:discussion} discusses strengths and limitations. Section~\ref{sec:related} reviews related work, and Section~\ref{sec:conclusion} concludes with future directions.

\section{Methodology}
\label{sec:methodology}
The methodlogies in this study includes the ontology design, data extraction, embeding, and the learning pipleline.

\subsection{Query-level Ontology}
As demonstrated in Fig~\ref{onto_db}, we name the objects and relations that matter to learned plan selection and make every run connected with the query, plan, and configurations. The figure is composed of nodes that contain two parts: the left side of the colon contains concepts, corresponding to the ontology TBox, and the right side contains entities, corresponding to the KG. 

The core classes include: \{Query, Plan, PlanNode, Table, Column, Index, Arm, Environment, Execution, Reward\}.
Typical relations are: (Query, hasPlan, Plan), (Plan, hasNode, PlanNode), (Execution, useArm, Arm). 

A PG extension instantiates the ABox per run (plan trees, per-operator stats, runtime/buffer metrics, GUC snapshots), so each execution ties a concrete (Query, Plan, Arm, Environment) to observed configurations and predictions.

\begin{figure}[!t]
	\centering
	\includegraphics[width=\columnwidth]{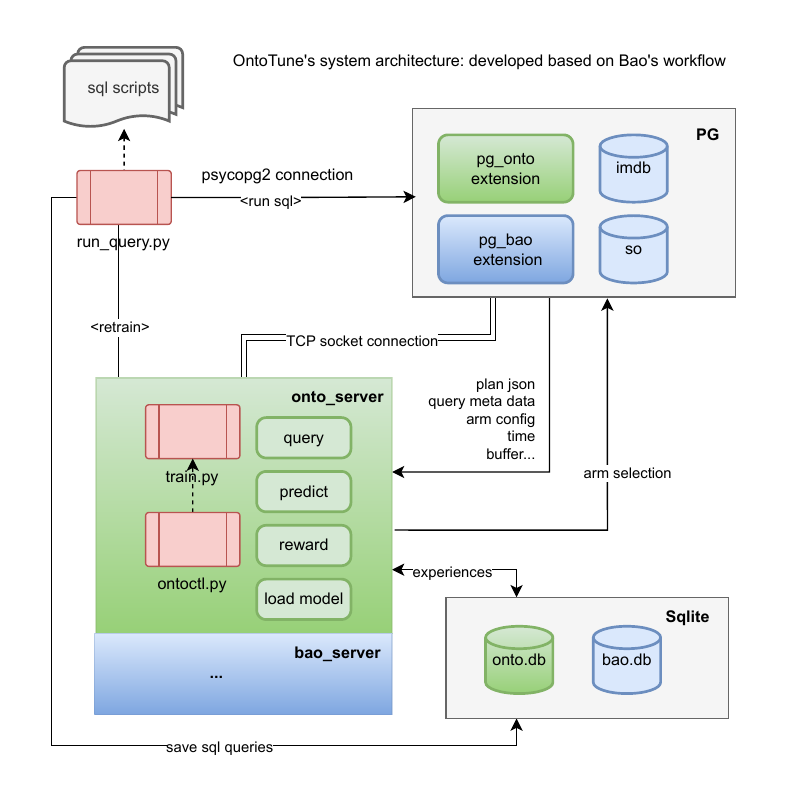}
	\caption{OntoTune's system architecture: developed based on Bao's workflow \cite{marcus2021bao}.}
	\label{fig:workflow}
\end{figure}

\begin{table*}[t]
	\caption{Representative features used in the \textsc{OntoTune} feature matrix.}
	\centering
	\small
	\setlength{\tabcolsep}{6pt}
	\renewcommand{\arraystretch}{1.12}
	\begin{tabularx}{\textwidth}{l L}
		\hline
		\textbf{Feature name} & \textbf{Meaning / How computed} \\
		\hline
		\texttt{tpl\_has\_distinct} & Query uses DISTINCT \\
		\texttt{tpl\_need\_sort\_for\_merge} & MergeJoin requires explicit sort (derived from plan/heuristics); Binary. \\
		\hline
		\texttt{tpl\_group\_by\_cols\_bucket\_\{0..3\}} & One-hot over bucketized \#GROUP BY columns; exactly one bucket set to 1 (others 0). \\
		\texttt{tpl\_rows\_bucket\_\{0..2\}} & One-hot over bucketized estimated result size;\\
		\hline
		\texttt{sql\_has\_window} & Window functions present; parsed into \texttt{template\_features} then broadcast. \\
		\texttt{sql\_has\_like} & LIKE patterns present; binary. \\
		\hline
		\texttt{sql\_num\_join\_bucket\_\{0..2\}} & One-hot over bucketized join count (e.g., 0, 1–2, $\ge$3). \\
		\texttt{sql\_num\_subquery\_bucket\_\{0..1\}} & One-hot over subquery count (0 vs.\ $\ge$1). \\
		\hline
		\texttt{plan\_cost\_share} & Per-table cost share normalized by total plan cost, then broadcast to that table’s columns. \\
		\texttt{plan\_rows\_share} & Per-table rows share normalized by total estimated rows, broadcast similarly. \\
		\hline
		\texttt{col\_cost\_from\_scan\_share} & Fraction of total plan cost attributed to scans mapped onto each referenced column. \\
		\texttt{col\_cost\_from\_agg\_share} & Fraction of cost from Aggregate nodes assigned to their argument/output columns. \\
		\hline
		\texttt{plan\_cost\_op\_HashJoin\_share} & Total cost share of HashJoin nodes over the whole plan; normalized to $[0,1]$. \\
		\texttt{plan\_cost\_op\_Sort\_share} & Total cost share of Sort nodes; normalized to $[0,1]$. \\
		\hline
	\end{tabularx}
	\label{tab:ontotune-features-sample}
\end{table*}

\subsection{Embedding}
\label{subsec:embedding}
For a context (template) \(\tau\), let the column universe \(\mathcal{U}_\tau\) enumerate referenced \texttt{Table}/\texttt{Column} instances. We build a matrix
\[
X_\tau \in \mathbb{R}^{R\times C},\quad C = |\mathcal{U}_\tau|,
\]
where rows are feature channels drawn from: (i) template/SQL indicators and buckets; (ii) plan-level cost/row shares; (iii) operator-type cost shares aggregated to referenced columns; and (iv) column traits (numeric/indexed/in-where/in-join/in-orderby).
Broadcasting from plan nodes \(p\) to columns \(u\in\mathcal{U}_\tau\) follows:
\begin{equation}
	\begin{aligned}
		\mathrm{plan\_cost\_share}(u)
		&= \sum_{p:\,u\in S(p)} \frac{\mathrm{cost}(p)}{\sum_{q} \mathrm{cost}(q)},\\
		\mathrm{plan\_rows\_share}(u)
		&= \sum_{p:\,u\in S(p)} \frac{\mathrm{rows}(p)}{\sum_{q} \mathrm{rows}(q)}.
	\end{aligned}
	\label{eq:plan-shares}
\end{equation}

We reform the data from SQL/plan/context to well-typed properties that are broadcasted into the matrix embedding and the epresentative properties can be found in Table~\ref{tab:ontotune-features-sample}. The same extraction pipeline can export adjacency/incidence for Tree/Graph models.

\begin{figure*}[htbp]
	\centering
	\includegraphics[width=0.9\linewidth]{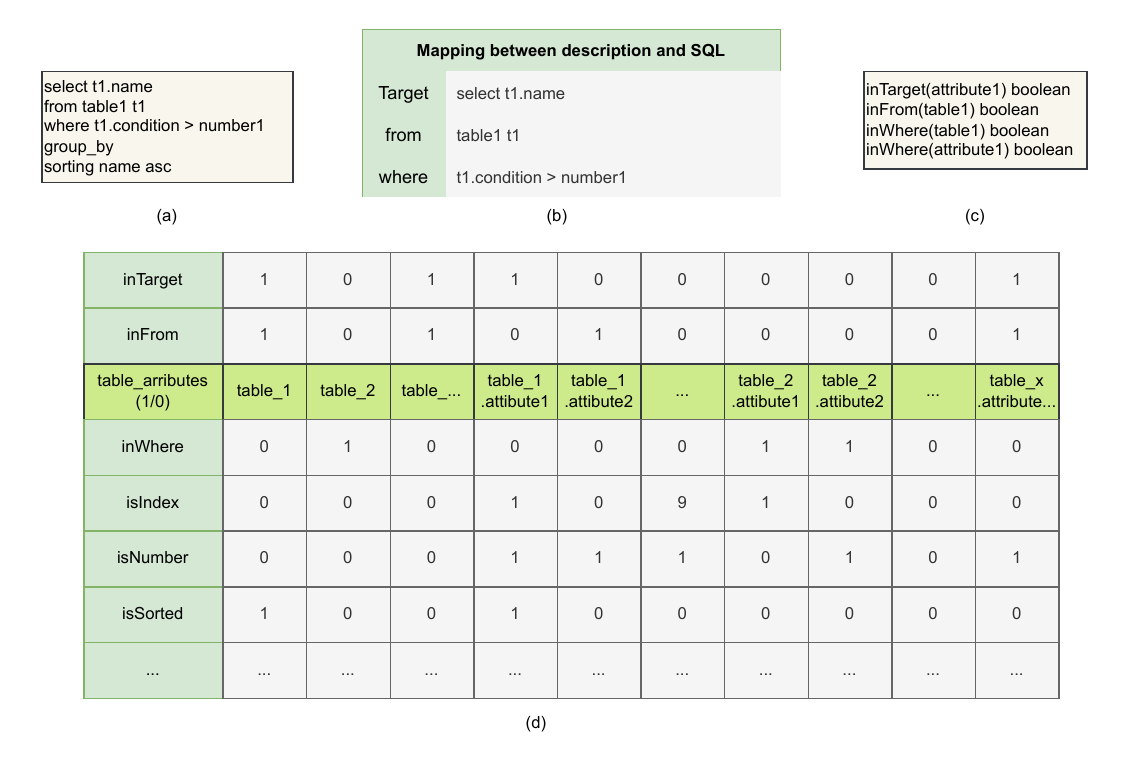}
	\caption{Partial Embedding of the OntoTune Ontology: for an SQL query}
	\label{fig:embedding}
\end{figure*}

The basic SQL-based ontology information is passed from the PG extension along with the plan, including analysis of a particular table attribute, such as whether it is numeric, whether it forms an index, and whether it needs sorting. When the values of those attributes are synthesized to the table level, they will decide the table's state in the matrix, as shown in Fig~\ref{fig:embedding} (d).

\subsection{OntoTune's System Architecture}
As shown in Fig~\ref{fig:workflow}, we use PG 12 as the basis for prototyping the OntoTune, as it is easier to make comparisons with the prior research work. We use C to implement the extension. We put the extraction of the DB statistics on the extension side, including the information from pg\_statistics, pg\_class, and pg\_namespace, as well as getting the information from GUC. The cost information is from the plans that are also sent from the extension side.

Before the execution, we store the SQLs in SQLite3, which are later used to extract extra query-level ontology information using SQLGLOT,  a Python library for parsing SQL statements. The "run\_query" script is to run the batches of queries and invoke the training between two batches. OntoServer provides the remote procedures, including recording the rewards of each query and making predictions for a specific query. 

\section{Implementation}
\label{sec:implementation}
To adopt CNN into the online reinforcement learing process, it is important to tackle a challenge that, when meeting with less samples, for example, for arm 1, which normally performs the worst, it won't be selected many times in the cold start running set, where an arm selection is decided by the PG's cost estimation, thus, in the second round, it tends to predict it to be the best arm to use.

We implemented a method to avoid this case (although for some cases, it is worth the exploration with a tailored design): by having an inverse of the predicted value – turn cost as reward and objective after training turns to getting the most significant reward, and finally, make a reverse to get the cost. This way, this process walks around the effect that the trained model gives the unseen arms the smallest cost, because the model after the inverse will just get the largest number to be the fastest (after 1-x), and the smallest number from the unseen arms is dropped. 

We consider $K$ candidate arms $a\in\{0,\dots,K-1\}$ for each incoming query under context $\tau$ with features $X_\tau$. Let $y_{\tau,a}>0$ denote the observed runtime cost (e.g., milliseconds) when arm $a$ is executed in context $\tau$. To stabilize the long-tailed distribution of runtimes, we apply a log--min--max scaler. We regress a \emph{high-is-good} reward and select arms by minimizing the recovered cost with optimism, candidate filtering and gated exploration.

\paragraph{Scaling and complement.}
Let \(y_{\tau,a}>0\) be runtime. Apply log–min–max scaling
\begin{align}
	\label{eq:phi_fwd}
	\phi(y) &= \frac{\log(1+y)-\ell_{\min}}{\ell_{\max}-\ell_{\min}} \in [0,1],\\
	\label{eq:phi_inv}
	\phi^{-1}(c) &= \exp\!\big(c(\ell_{\max}-\ell_{\min})+\ell_{\min}\big) - 1 .
\end{align}

\paragraph{Reward--Cost Complement Transform}
Define the complement
\begin{equation}
	\label{eq:J}
	J(x)=1-x,
\end{equation}
and the (scaled) reward
\begin{equation}
	\label{eq:r}
	r_{\tau,a}=J\!\big(\phi(y_{\tau,a})\big)=1-\phi(y_{\tau,a})\in[0,1].
\end{equation}

\paragraph{Model and loss.}
A CNN \(f_\theta\) predicts reward from \((X_\tau,a)\):
\begin{equation}
	\label{eq:pred}
	\hat r_{\tau,a}=f_\theta(X_\tau,a)\in[0,1].
\end{equation}
Recover scaled/real cost for selection:
\begin{equation}
	\label{eq:invert}
	\hat c_{\tau,a}=J(\hat r_{\tau,a})=1-\hat r_{\tau,a},\qquad
	\tilde y_{\tau,a}=\phi^{-1}(\hat c_{\tau,a}).
\end{equation}
Train with MSE in reward space:
\begin{equation}
	\label{eq:mse}
	\mathcal{L}(\theta)=\frac{1}{|D|}\sum_{(\tau,a)\in D}\big(f_\theta(X_\tau,a)-r_{\tau,a}\big)^2.
\end{equation}

\paragraph{Scoring and candidate filtering.}
With per-context counts \(n_{\tau,a}\) and optimism hyperparameter \(\beta\ge 0\),
\begin{equation}
	\label{eq:score}
	s_{\tau,a}=\tilde y_{\tau,a}-\frac{\beta}{\sqrt{\max(1,n_{\tau,a})}}.
\end{equation}
Keep the best \(K_{\text{top}}\) arms after removing a banned tail \(B_\tau\) and enforcing a minimum sample threshold \(n_{\min}\):
\begin{equation}
	\label{eq:Acand}
	A^{\mathrm{cand}}_{\tau}=\Big(\mathrm{TopK}_a\big(s_{\tau,a};K_{\text{top}}\big)\setminus B_\tau\Big)\setminus\{a:n_{\tau,a}<n_{\min}\}.
\end{equation}

\paragraph{Gated $\varepsilon$-greedy and final decision.}
Let \(N_\tau\) be the frequency of \(\tau\), and \(\texttt{estC}_\tau\) a normalized batch cost:
\begin{equation}
	\label{eq:eps}
	\varepsilon_\tau=\max\!\Big(\varepsilon_{\min},\frac{\varepsilon_0}{\sqrt{N_\tau}}\Big)\cdot\frac{1}{1+\texttt{estC}_\tau}.
\end{equation}
Choose
\begin{equation}
	\label{eq:policy}
	a^\star_\tau=
	\begin{cases}
		\arg\min_{a\in A^{\mathrm{cand}}_{\tau}} s_{\tau,a}, & \text{with prob. } 1-\varepsilon_\tau,\\[2pt]
		\mathrm{Uniform}\!\left(A^{\mathrm{cand}}_{\tau}\right), & \text{with prob. } \varepsilon_\tau.
	\end{cases}
\end{equation}\\

\section{Case Study and Experiments}
\label{sec:experiments}
CNN is used to capture features from the matrix formed by the ontology embedding and to perform regression on which arm's plan will have the best execution time – reward prediction. The purpose of this experiment is to serve as a case study to demonstrate how the OntoTune platform can provide insights into what should be added or emphasized next to make the platform better suited to support query optimization research. 

The reported results are confirmed by three independent runs using the GitHub release (https://github.com/songhui01/OntoTune.git) and the stated configuration. Empirically, the model sometimes selects the best arms for long-tail queries in some datasets and configurations, but in another dataset, it can take longer due to selecting a very expensive arm. 

\subsection{Configurations}
Unless otherwise stated, all experiments were executed on Ubuntu 24.04 with PostgreSQL 12. The workload is the StackOverflow (so\_pg) dataset. The host machine is equipped with an Intel Core i5-class CPU and an NVIDIA Blackwell GPU (20 GB VRAM).

We fixed the following GUC settings to make results comparable across runs:
\texttt{shared\_buffers=2GB}, \texttt{work\_mem=4MB}, \texttt{maintenance\_work\_mem=64MB}, \texttt{effective\_io\_concurrency=2}, \texttt{max\_worker\_processes=8}, \texttt{max\_parallel\_workers\_per\_gather=1}, \texttt{max\_parallel\_workers=1}, and \texttt{effective\_cache\_size=2GB}.
All other parameters remained at PostgreSQL defaults.

\subsection{Results}
On StackOverflow (so\_pg) under the fixed configuration, OntoTune outperforms PostgreSQL for one representative random seed (Fig. 4): the orange curve rises more steeply and completes the 200-query batch several minutes earlier, with fewer long plateaus. For another seed (Fig. 5), OntoTune underperforms: after an initial rise, it exhibits extended stalls and finishes noticeably later than PostgreSQL. These trends were consistent across three independent repetitions for each seed.

\begin{figure}[!t]
	\centering
	\includegraphics[width=0.8\columnwidth]{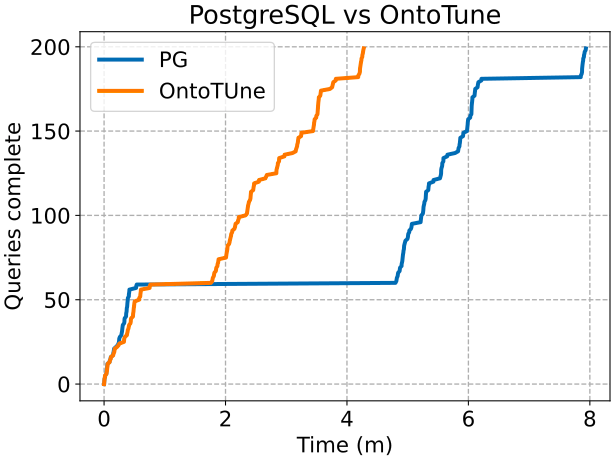}
	\caption{Queries vs Time for PostgreSQL and OntoTune with Random Set (42).}
	\label{fig:queries_vs_time_42}
\end{figure}

\begin{figure}[!t]
	\centering
	\includegraphics[width=0.8\columnwidth]{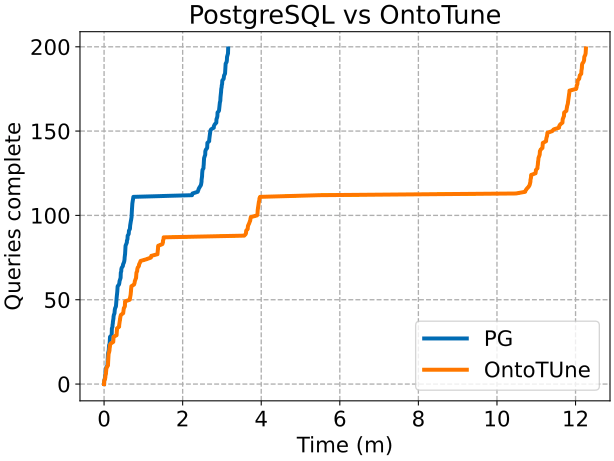} 
	\caption{Queries vs Time for PostgreSQL and OntoTune with Random Set (49).}
	\label{fig:queries_vs_time_49}
\end{figure}

\subsection{Analysis}
The win in Fig. 4 is consistent with our design: the learned policy leverages prior executions to identify better arms to shorten completion time by avoiding long stalls. The curve shows that the method performs well when meeting the two long-tailing queries. The loss in Fig. 5 occurs when early exploration or misranking admits an expensive arm into the candidate set, as a result, a few costly queries dominate wall-time and flatten the curve.

A plausible explanation for Fig.~5 is that the system was still in the learning phase; with sufficient time and samples, OntoTune is expected to recover and improve. To make such improvements reliable, we plan to enrich the ontology/knowledge graph with tail-aware nodes and relations (e.g., spill risk, operator-level memory pressure, join-selectivity drift) so that root causes of long tails can be systematically mitigated via arm choice or dynamic configuration~\cite{giannakouris2025lambda}.

\section{Discussion}
\label{sec:discussion}
Our study positions OntoTune as a platform rather than a single model: it encodes queries, plans, and execution context into an ontology/knowledge-graph layer and learns over this representation. This separation between the semantic layer and the learner enables CNNs (and, in principle, graph/TreeConv variants) to operate on a consistent, query-level abstraction. Compared with Bao—which learns features directly on the plan tree—our approach explicitly transforms SQL/plan/context signals into ontology relations and features. The case study shows feasibility: a convolutional regressor can select effective arms and, in favorable settings, avoid bad plans that produce long stalls.

One limitation of the study is that the performance can be sensitive to environment and parameter changes (e.g., GUCs, statistics refresh), leading to instability even when the selected arms are usually reasonable. The ontology currently covers most—but not all—categories we need; tail-aware attributes are still incomplete. Because learning requires many batches of executed queries, our experiments used relatively small subsets, and long-horizon benefits were not thoroughly tested. These are threats to validity and define the boundaries of what the present evidence supports.

In the long run, instability can be mitigated with environment-aware transfer learning (offline pretraining followed by light online adaptation), and to generalize across workloads by mining templates and using ontology-graph similarity for retrieval and warm starts. On the representation side, the TBox/ABox can be broadened with tail-focused nodes and relations, then instrument a closed loop that (i) automatically updates the knowledge base from experimental feedback, (ii) schedules/evaluates new runs, and (iii) deploys safety fallbacks when predicted risk or uncertainty is high. Together, these steps aim to make ontology-driven learning a robust and systematic path toward practical query optimization.

\section{Related Work}
\label{sec:related}
Learned query optimization has progressed from plan-tree encoders to end-to-end selection policies. Bao learns over operator trees and uses a tree-convolutional network to rank alternative knobs/arms, demonstrating practical gains under controlled settings \cite{marcus2021bao}. Subsequent systems extend this direction with alternative learners and search strategies (e.g., learned re-ranking and tighter candidate exploration). Lero further explores model-driven selection by embedding query/plan signals and learning when a nondefault choice yields lower latency \cite{zhu2023lero}. These works share the premise that a learned model can exploit execution history to guide per-query decisions.

Our work differs in representation: instead of encoding the physical plan solely as a tree, we explicitly transform SQL/plan/context information into an ontology/knowledge-graph and derive a matrix (and optional graph) embedding from that semantic layer. This design separates the what (semantics and provenance) from the how (choice of learner), allowing CNNs—as shown in our case study—and, in future, GCN/GAT or TreeConv, to operate on one consistent, query-level abstraction.

Orthogonal to the above, LLM-based tuning has been explored for database configuration and query improvement, e.g., Lambda-Tune \cite{giannakouris2025lambda} and GPTuner \cite{lao2025gptuner}. These methods focus on generating or selecting configurations and often do not expose full metrics/provenance for ablation. Our platform is complementary: the ontology captures relations among queries, plans, environments, and outcomes, enabling controlled ablations and explanations (e.g., which relations/features correlate with long-tail stalls). Pairing LLMs with ontology/KG for template discovery in this study can be a natural next step.

\section{Conclusion}
\label{sec:conclusion}
We presented OntoTune, an ontology-driven platform that records queries, plans, arms, environments, and outcomes as first-class objects and turns them into a reproducible knowledge base. From this layer, we derive a consistent embedding—matrix features (and an optional adjacency matrix) suitable for convolutional and graph models. A CNN case study demonstrated the end-to-end workflow, including data pipeline and arm selection, and showed that the learned policy can leverage prior executions to avoid disk- or memory-heavy plans and improve completion time under a fixed PostgreSQL configuration.

This work is a first step. The current ontology is intentionally minimal and results can still be sensitive to environment and parameter changes. Going forward, we will expand tail-aware features in the ontology/KG, apply environment-aware transfer learning for stability, and explore GCN/GAT/TreeConv as complementary learners. We also plan to pair the ontology with LLM-assisted template discovery and configuration suggestions, and to close the loop by automatically updating the knowledge base from feedback and scheduling new runs. The goal of this study is a systematic, explainable, and reproducible path to learned query optimization.

% ---- Bibliography ----
\bibliographystyle{IEEEtran}
\bibliography{refs}

@inproceedings{marcus2021bao,
	title={Bao: Making learned query optimization practical},
	author={Marcus, Ryan and Negi, Parimarjan and Mao, Hongzi and Tatbul, Nesime and Alizadeh, Mohammad and Kraska, Tim},
	booktitle={Proceedings of the 2021 International Conference on Management of Data},
	pages={1275--1288},
	year={2021}
}

@article{zhu2023lero,
	title={Lero: A learning-to-rank query optimizer},
	author={Zhu, Rong and Chen, Wei and Ding, Bolin and Chen, Xingguang and Pfadler, Andreas and Wu, Ziniu and Zhou, Jingren},
	journal={arXiv preprint arXiv:2302.06873},
	year={2023}
}

@inproceedings{ding2019ai,
	title={Ai meets ai: Leveraging query executions to improve index recommendations},
	author={Ding, Bailu and Das, Sudipto and Marcus, Ryan and Wu, Wentao and Chaudhuri, Surajit and Narasayya, Vivek R},
	booktitle={Proceedings of the 2019 International Conference on Management of Data},
	pages={1241--1258},
	year={2019}
}

@inproceedings{ma2020active,
	title={Active learning for ML enhanced database systems},
	author={Ma, Lin and Ding, Bailu and Das, Sudipto and Swaminathan, Adith},
	booktitle={Proceedings of the 2020 ACM SIGMOD International Conference on Management of Data},
	pages={175--191},
	year={2020}
}

@article{giannakouris2025lambda,
	title={$\lambda$-tune: Harnessing large language models for automated database system tuning},
	author={Giannakouris, Victor and Trummer, Immanuel},
	journal={Proceedings of the ACM on Management of Data},
	volume={3},
	number={1},
	pages={1--26},
	year={2025},
	publisher={ACM New York, NY, USA}
}

@article{gadde2024intelligent,
	title={Intelligent Query Optimization: AI Approaches in Distributed Databases},
	author={Gadde, Hemanth},
	journal={International Journal of Advanced Engineering Technologies and Innovations},
	volume={1},
	number={2},
	pages={650--691},
	year={2024}
}

@article{heitz2019join,
	title={Join query optimization with deep reinforcement learning algorithms},
	author={Heitz, Jonas and Stockinger, Kurt},
	journal={arXiv preprint arXiv:1911.11689},
	year={2019}
}

@inproceedings{ortiz2018learning,
	title={Learning state representations for query optimization with deep reinforcement learning},
	author={Ortiz, Jennifer and Balazinska, Magdalena and Gehrke, Johannes and Keerthi, S Sathiya},
	booktitle={Proceedings of the Second Workshop on Data Management for End-To-End Machine Learning},
	pages={1--4},
	year={2018}
}

@inproceedings{lee2023relation,
	title={Relation Modeling on Knowledge Graph for Interoperability in Recommender Systems},
	author={Lee, Seungjoo and Ahn, Seokho and Seo, YoungDuk},
	booktitle={Proceedings of the 38th ACM/SIGAPP Symposium on Applied Computing},
	pages={751--758},
	year={2023}
}

@article{rajabi2024knowledge,
	title={Knowledge-graph-based explainable AI: A systematic review},
	author={Rajabi, Enayat and Etminani, Kobra},
	journal={Journal of information science},
	volume={50},
	number={4},
	pages={1019--1029},
	year={2024},
	publisher={Sage Publications Sage UK: London, England}
}

@article{ben2021novel,
	title={A novel approach for learning ontology from relational database: from the construction to the evaluation},
	author={Ben Mahria, Bilal and Chaker, Ilham and Zahi, Azeddine},
	journal={Journal of Big Data},
	volume={8},
	number={1},
	pages={25},
	year={2021},
	publisher={Springer}
}

@article{simsek2023knowledge,
	title={A knowledge graph perspective on knowledge engineering},
	author={Simsek, Umutcan and K{\"a}rle, Elias and Angele, Kevin and Huaman, Elwin and Opdenplatz, Juliette and Sommer, Dennis and Umbrich, J{\"u}rgen and Fensel, Dieter},
	journal={SN Computer Science},
	volume={4},
	number={1},
	pages={16},
	year={2023},
	publisher={Springer Nature BV}
}

@article{zhao2022queryformer,
	title={Queryformer: A tree transformer model for query plan representation},
	author={Zhao, Yue and Cong, Gao and Shi, Jiachen and Miao, Chunyan},
	journal={Proceedings of the VLDB Endowment},
	volume={15},
	number={8},
	pages={1658--1670},
	year={2022},
	publisher={VLDB Endowment}
}

@phdthesis{yang2022machine,
	title={Machine learning for query optimization},
	author={Yang, Zongheng},
	year={2022},
	school={University of California, Berkeley}
}

@article{zhang2019graph,
	title={Graph convolutional networks: a comprehensive review},
	author={Zhang, Si and Tong, Hanghang and Xu, Jiejun and Maciejewski, Ross},
	journal={Computational Social Networks},
	volume={6},
	number={1},
	pages={1--23},
	year={2019},
	publisher={Springer}
}

@article{velivckovic2017graph,
	title={Graph attention networks},
	author={Veli{\v{c}}kovi{\'c}, Petar and Cucurull, Guillem and Casanova, Arantxa and Romero, Adriana and Lio, Pietro and Bengio, Yoshua},
	journal={arXiv preprint arXiv:1710.10903},
	year={2017}
}

@article{ma2022understanding,
	title={Understanding of Convolutional Neural Network (CNN): A Review.},
	author={Ma'arif, Alfian and Rahmaniar, Wahyu and Fathurrahman, Haris Imam Karim and Frisky, Aufaclav Zatu Kusuma and others},
	journal={International Journal of Robotics \& Control Systems},
	volume={2},
	number={4},
	year={2022}
}

@article{leis2015good,
	title={How good are query optimizers, really?},
	author={Leis, Viktor and Gubichev, Andrey and Mirchev, Atanas and Boncz, Peter and Kemper, Alfons and Neumann, Thomas},
	journal={Proceedings of the VLDB Endowment},
	volume={9},
	number={3},
	pages={204--215},
	year={2015},
	publisher={VLDB Endowment}
}

@article{lao2025gptuner,
	title={GPTuner: An LLM-Based Database Tuning System},
	author={Lao, Jiale and Wang, Yibo and Li, Yufei and Wang, Jianping and Zhang, Yunjia and Cheng, Zhiyuan and Chen, Wanghu and Tang, Mingjie and Wang, Jianguo},
	journal={ACM SIGMOD Record},
	volume={54},
	number={1},
	pages={101--110},
	year={2025},
	publisher={ACM New York, NY, USA}
}

@misc{marcus_stack_sopg,
	author       = {Ryan Marcus},
	title        = {Stack (so\_pg) Dataset for PostgreSQL},
	year         = {2021},
	howpublished = {\url{https://rmarcus.info/stack.html}},
	note         = {PostgreSQL 12/13 archives and 5000+ queries}
}

\end{document}